# In-situ growth of superconducting MgB$_2$ thin films by molecular beam epitaxy

K Ueda[*], M. Naito

NTT Basic Research Laboratories, NTT Corporation,

3-1 Wakamiya, Morinosato, Atsugi, Kanagawa 243-0198, Japan

**Abstract**

The in-situ growth of superconducting MgB$_2$ thin films was examined from various perspectives. The paper discusses (1) growth temperature, (2) the effect of excess Mg, (3) the effect of residual gas during growth, (4) the effect of in-situ annealing, (5) thickness dependence and (6) the effect of substrates. Our results provide a guide to the preparation of high-quality superconducting MgB$_2$ films for potential electronics applications.

## 1. Introduction

The discovery of superconductivity at 39 K in $MgB_2$ [1] has generated great scientific and technological interest. $MgB_2$ has many properties that make it very attractive for superconducting electronics. Unlike high-$T_c$ cuprates, $MgB_2$ may be suitable, despite its lower $T_c$, for fabricating good Josephson junctions. This is because it has less anisotropy, fewer material complexities, fewer interface problems, and a longer coherence length ($\xi \sim 5$ nm). The reliable fabrication of Josephson junctions requires high-quality thin films of $MgB_2$. In this article, we report our in-situ growth of $MgB_2$ films by coevaporation.

There have already been many reports on the synthesis of superconducting $MgB_2$ films [2-14]. There are two complicating problems related to the preparation of superconducting $MgB_2$ films. These are the high sensitivity of Mg to oxidation and the high Mg vapor pressure required for the thermodynamic stability of the superconducting $MgB_2$ phase. The former problem can be avoided by depositing $MgB_2$ films either in an ultra-high vacuum or in a reducing atmosphere containing hydrogen. The latter problem is more serious, and there are two ways to overcome it. One is two-step synthesis, in which amorphous B (or Mg-B composite) precursors are *ex-situ* annealed at high temperatures with high Mg vapor pressure usually in a confined container [2-6]. This two-step process produces good crystalline films with good superconducting properties although it may not be favorable for multilayer device fabrication. The other is *in-situ* (as-grown) synthesis, in which films are grown *in-situ* at low temperatures of around 300°C at the relatively low Mg vapor pressure ($10^{-5} \sim 10^{-6}$ Torr) that is compatible with many vacuum deposition techniques. *In-situ* synthesis makes multilayer deposition feasible although in most cases it only produces poor crystalline films with a $T_c$ of ~ 35 K, which is slightly below the bulk value [10-13]. This is because good epitaxial growth is impeded due to the limited growth temperatures below ~300°C. By employing a special combination of physical and chemical vapor deposition techniques, Zeng *et al.* have very recently achieved the *in-situ* epitaxial growth of $MgB_2$ films at around 750°C at a high Mg vapor pressure (10-100 mTorr) [14]. However, it remains to be seen whether this method is as suitable for multilayer deposition as conventional physical vapor deposition.

In the last 18 months since the discovery of $MgB_2$, we have performed a fairly comprehensive survey of the parameters in *in-situ* $MgB_2$ film growth by coevaporation. In this article, we report the results of this survey. The following issues will be discussed [10].

(1) Growth temperature
(2) Effect of excess Mg
(3) Effect of residual gas during growth
(4) Effect of in-situ annealing
(5) Thickness dependence
(6) Effect of substrates

## 2. Experimental

We grew $MgB_2$ thin films in a customer-designed MBE chamber (basal pressure < $10^{-9}$ Torr) from pure metal sources using multiple electron guns [15, 16]. The evaporation beam flux of each element was controlled by electron impact emission spectrometry (EIES) via feedback loops to the electron guns. The flux ratio of Mg to

B was changed so that it was from 1 to 10 times as high as the nominal ratio. The growth rate was 1.5 - 2 Å/s, and the film thickness was 1000 Å for typical films. We used sapphire C, sapphire R, H-terminated Si (111), $SrTiO_3$ (100), and glass substrates (Corning #7059).

We characterized the structure and crystallinity by X-ray diffraction (XRD, $2\theta - \theta$ scan) and reflection high-energy electron diffraction (RHEED). We measured resistivity by the standard four-probe method using electrodes formed by Ag evaporation. The composition was determined by inductively coupled plasma (ICP) spectrometric analysis. We sometimes found that ICP analysis underestimated the B content of the films perhaps due to the low solubility of B in usual acids. These questionable data were not included in the results described below.

## 3. Results and Discussion
### 3.1 Growth temperature

A very high Mg vapor pressure is required when depositing $MgB_2$ films at elevated temperatures. In other words, the growth temperature when depositing $MgB_2$ films has to be low at a low Mg vapor pressure. This conclusion can be reached both from the thermodynamic stability of the $MgB_2$ phase and from the growth kinetics. Given an Mg vapor pressure of $10^{-4}$ Torr, which is the upper limit compatible with vacuum evaporation techniques, the thermodynamic calculation for the Mg-B binary phase diagram reported by Liu et al [17], as shown in Fig. 1, predicts that the superconducting $MgB_2$ phase will decompose above 535°C. On the other hand, the growth kinetics on a substrate or film surface predicts that non-bonding Mg re-evaporates from the surface for a growth temperature above ~ 300°C in $10^{-4}$ Torr of Mg vapor according to the equilibrium Mg vapor pressure curve [18]. Both predictions suggest that the upper limit of the substrate temperature for $MgB_2$ growth by coevaporation will be 300 ~ 400°C.

Consistent with these considerations, we actually observed a significant Mg loss in films grown above a substrate temperature ($T_s$) of 300°C. Figure 2(a) shows the molar ratio of Mg to $B_2$ evaluated by ICP analysis in films grown at temperatures of 200°C to 500°C and also with Mg rates 1.3 to 10 times the nominal rate. The films grown above 300°C were significantly deficient in Mg even with a 10 times higher Mg rate. These Mg deficient films were transparent and insulating. This indicates that the growth temperature limit is 300°C for the in-situ growth of superconducting $MgB_2$ films by coevaporation. However, a slight and uncontrollable change (5°C) in the growth temperature near $T_s$ = 300°C leads to a dramatic change in composition, and makes the results irreproducible. We have therefore chosen the following as our typical growth conditions: $T_s$ = 280°C with an Mg rate maintained at three times the nominal rate. The resultant as-grown films on sapphire C substrates typically exhibited XRD patterns and ρ-T curves as shown in Fig. 3. The films had a c-axis preferred orientation, and their $T_c$ and resistivity were 32 - 35 K and 30 – 50 μΩcm at 300 K, respectively, which should be compared with the bulk single crystal values of 39 K and 5-10 μΩcm.

When the growth temperature is lowered, the crystallinity becomes poor judging from the XRD peak intensities, and simultaneously the superconducting properties become degraded as shown in Fig. 2(b). This figure shows the superconducting transition temperature of $MgB_2$ films grown at different $T_s$ values on sapphire C with an Mg rate of twice the nominal rate. The $T_c$ value decreases monotonically as $T_s$ decreases as long as the $MgB_2$ phase is formed, and disappears below ~150°C. This $T_c$ suppression is not only due to poorer crystallinity but also due to excess Mg as described below.

3.2 Effect of excess Mg

Figure 4 shows the superconducting properties of $MgB_2$ films grown on sapphire C at a fixed substrate temperature (280°C) but with different Mg rates (1 ~ 10 times). The desired phase is not formed with Mg rates of less than 3 times the nominal rate. At higher Mg rates, $T_c$ gradually decreases and eventually becomes around 15 K at 10 times the nominal Mg rate. A reduction in resistivity accompanies this $T_c$ suppression, indicating the presence of metal Mg in these films. This result demonstrates the harmful effect of excess Mg on the superconducting properties, which is either due to the proximity effect with normal Mg metal or the formation of nonstoichiometric $Mg_{1+x}B_2$ that is hitherto unknown to be formed in bulk synthesis.

3.3 Effect of residual gas during growth

In our early attempts to grow $MgB_2$, we had some difficulty in obtaining reproducible results. For example, the resistivity of Fig. 4(b) is considerably higher than the typical value in Fig. 3(b). These irreproducible results can be partly explained by the fact that the sticking coefficient of Mg varies dramatically near $T_s$ ~ 300°C. However, we have noticed another reason for the irreproducibility, namely the effect of residual oxygen on $MgB_2$ film growth. In fact, we obtained the high-resistivity film shown in Fig. 4(b) when we had a tiny leak in our MBE chamber, which gave an oxygen partial pressure of $P_{O2}$ ~ $4\times10^{-9}$ Torr. Based on this experience, we undertook a systematic investigation of the effect of residual gas on the growth of $MgB_2$ films. We studied the effects of hydrogen and nitrogen in addition to that of oxygen. In this experiment, molecular $O_2$, $H_2$ or $N_2$ gas was introduced into the chamber during growth. We varied the gas pressure from $3\times10^{-9}$ to $8\times10^{-6}$ Torr.

First, we describe the effect of oxygen on $MgB_2$ growth. Figure 5 shows the XRD patterns and r-T curves of $MgB_2$ films grown on sapphire C under two different partial oxygen pressures. Weak but definite $MgB_2$ (00$l$) peaks are observed for the film grown in $P_{O2}$ < $1.0\times10^{-10}$ Torr whereas no peak is observed for the film grown in $P_{O2}$ ~ $3.7\times10^{-9}$ Torr. The resistivity of the latter film is much higher although $T_c$ is not greatly suppressed (only ~ 3 K). For $P_{O2}$ > $10^{-7}$ Torr, the films were transparent and insulating, indicating that no $MgB_2$ phase was formed. These observations strongly indicate that residual oxygen is very harmful to $MgB_2$ film growth, even with $P_{O2}$ as low as $1\times10^{-9}$ Torr. This result can be readily understood from the high sensitivity of Mg to oxidation.

In contrast to oxygen, residual hydrogen and nitrogen have a negligible or even a slightly favorable effect on $MgB_2$ film growth. Figure 6 shows superconducting transitions of $MgB_2$ films grown on sapphire C in $H_2$ or $N_2$ gas with different partial pressures. The introduction of $H_2$ and $N_2$ gas even improved $T_c$ by about 1 K. The maximum $T_c$ improvement (~1.5 K) was achieved by the introduction of $N_2$ gas at $3.8\times10^{-6}$ Torr. The significant resistivity increase with a hydrogen partial pressure of $5\times10^{-6}$ Torr could be due to the fact that our hydrogen gas was less pure than the nitrogen gas. Figure 7 shows a plot of the superconducting transition temperature as a function of $P_{H2}$ or $P_{N2}$. When $N_2$ gas was introduced, $T_c$ was at its maximum around $P_{N2}$ ~ $4\times10^{-6}$ Torr. We repeated the experiment to confirm that this effect really exists rather than being the results of scattering, and also obtained a similar trend in the second round ($N_2$-2).

The origin of the slight improvement in $T_c$ with residual $H_2$ or $N_2$ is unclear. It is conceivable, however, that hydrogen may prevent Mg from oxidation. With nitrogen, the origin may be more complicated. One hint is provided by the fact that film crystallinity is somewhat improved by the introduction of nitrogen as shown in Fig. 8. This improved crystallinity is seemingly reflected in the lower resistivity of these films. Nitrogen may play a catalytic role in promoting the chemical reaction (Mg + 2B -> $MgB_2$) and crystallization. In this connection, it should be noted that Lee et al. have succeeded in growing $MgB_2$ single crystals in the Mg-B-N ternary system [19].

3.4 Effect of in-situ annealing

As seen from the above results, the properties of as-grown films are inferior to those of bulk single crystals or *ex-situ* post-annealed films in that the superconducting transition temperature is lower, namely $T_c$(end) is ~ 35 K at its highest, and the resistivity has a higher value and a weaker temperature dependence. With the aim of improving the superconducting properties of $MgB_2$ films, we attempted in-situ post-annealing for as-grown films. The in-situ post-annealing was performed just after growth at annealing temperatures ($T_a$) of 380 ~ 680°C for 10 minutes while exposing the films to Mg flux (~ 4.5 Å/sec). From the thermodynamic viewpoint, our recipe of the in-situ post-annealing may be outside the stability field of the $MgB_2$ phase as shown below. However, the very slow kinetics of the decomposition process predicts that the inside of the films will be protected from decomposition (see Fig. 1) [18]. In fact, the in-situ post-annealing slightly improved the inside of the films. Figure 9 provides a summary of the results. The XRD results suggest that annealing at $T_a$ below 503°C did not improve the crystallinity of the films whereas annealing at $T_a$ = 526°C increased the peak intensities significantly (Fig. 9(a)). The RHEED observations made during annealing indicated that the surface of the films had already started to decompose at $T_a$ ~ 500°C. The RHEED patterns became halo-like with annealing at $T_a$ ~ 550°C, above which the inside of the films also started to decompose. This situation did not change greatly even when the Mg flux rate was tripled to ~ 15.0 Å/sec.

For $T_a$ below 503°C, although there was no improvement in the crystallinity as mentioned above, the superconducting transition temperature increased gradually with increasing $T_a$, and a maximum improvement of ~ 2 K was achieved by annealing at $T_a$ = 503°C. The resistivity was also improved although there was some scattering in the data. This improved superconductivity may result from a partial elimination of an unfavorable feature in the grain boundaries. The $T_c$ of the film annealed at $T_a$ = 526°C was slightly lower than that of the film annealed at $T_a$ = 503°C in spite of its better crystallinity.

Fig. 10 compares AFM images (1 μm×1 μm view) of the as-grown and in-situ post-annealed films. The grain size was almost the same in the as-grown films and the in-situ annealed films at 480°C (200 - 400 Å), namely in-situ annealing at 480°C does not increase the grain size. However, the grain size is slightly larger in the films annealed at 526°C. These AFM results seem to be consistent with the above XRD results (Fig. 9(a)).

The surfaces of all the films were fairly smooth, and the root-mean-square roughness ($R_{MS}$) and the average roughness ($R_a$) were 22.3 Å and 17.7Å for the as-grown films, 17.4 Å and 13.7Å for the post-annealed films at 480°C, and 29.6Å and 23.1Å for the post-annealed films at 526°C, respectively. The $R_{MS}$ and $R_a$ of the film post-annealed at 526°C were the largest reflecting the increase in the grain size.

3.5 Thickness dependence

The thickness dependence of the superconductivity was examined for both of as-grown and post-annealed films. We undertook this investigation specifically for the purpose of evaluating the degraded surface thickness of post-annealed films. The film thickness, which was controlled by changing the deposition time, was varied from 25 to 1000 Å. In this experiment, post-annealed films were prepared at $T_a$ = 480°C. The thickness dependences of the r-T curves are shown for as-grown films in Fig. 11, and for post-annealed films in Fig. 12. In both cases, $T_c$ decreased and the resistivity increased with decreasing film thickness. Figure 13 is a plot of $T_c$ as a function of film thickness. For film thicknesses greater than 100 Å, the $T_c$ of the post-annealed films is higher than that of the as-grown films, but the situation is reversed at 50 Å. As-grown film, even with a thickness of 50 Å, is superconducting with a $T_c$(end) of ~ 10 K whereas the post-annealed film with a thickness of 50 Å is insulating. Figure 14 compares the RHEED patterns of the as-grown and post-annealed 50 Å films. The spotty patterns were changed to halo patterns by annealing [20]. These observations indicate surface decomposition due to Mg loss from the surface, which implies that our in-situ post-annealing conditions even with $T_a$ = 480°C are out of the thermodynamic stability of the $MgB_2$ phase. Our recent tunneling experiments support this conclusion: post-annealed films have a dead surface (nonsuperconducting) layer and are unsuitable for fabricating junctions [21].

3.6 Effect of substrates

Finally we describe the effect of the substrates. Since our as-grown films are not highly crystalline, the epitaxial effect of the substrates is rather unimportant as compared with high-$T_c$ cuprate films, in which the substrates play a significant role in improving the film quality [22]. However, we observed that $MgB_2$ films have a weak but finite substrate dependence. Figure 15 shows the superconducting transitions of $MgB_2$ films on different substrates: (a) as-grown and (b) post-annealed at $T_a$ = 480°C. In the as-grown films, $T_c^{zero}$ was 32.2, 33.2, 30.3 and 29.5 K for Si (111), sapphire C, sapphire R and $SrTiO_3$ (100), respectively. With the post-annealed films, $T_c^{zero}$ was 36.8, 36.6, 36.1 and 35.7 K for Si (111), sapphire C, sapphire R and $SrTiO_3$ (100), respectively. In both the as-grown and post-annealed films, $T_c^{zero}$ was slightly higher on Si (111) and sapphire C than on sapphire R and $SrTiO_3$ (100). Si (111) and sapphire C have a hexagonal surface whereas sapphire R and $SrTiO_3$ (100) have a square or rectangular surface. $MgB_2$ has a hexagonal crystal structure with hexagonal Mg and B planes stacked alternately along the c axis [1]. Therefore it is understandable that substrates with a hexagonal surface provide slightly better results.

We also prepared $MgB_2$ films on glass substrates (Corning #7059). The XRD pattern and the r-T curve of the film grown at $T_s$ = 280°C are shown in Fig. 16. The film showed a $T_c^{zero}$ of 33.4 K although no distinct peaks were observed in the XRD pattern. The RHEED pattern contains rings in addition to the spots that are commonly observed on crystalline substrates. This result demonstrates that fair quality $MgB_2$ films can be grown even on amorphous substrates such as glass.

**4. Summary**

We examined various aspects of the in-situ growth of superconducting $MgB_2$ thin films. The following summarizes each subsection of this article.

(1) Growth temperature: The growth temperature for depositing $MgB_2$ films has to be low (~300°C) at a low Mg vapor pressure (~$10^{-6}$ Torr). This conclusion can be reached both from the thermodynamics of the stability of the $MgB_2$ phase and also from the growth kinetics.

(2) Effect of excess Mg: Excess Mg is harmful to the superconducting properties of $MgB_2$ due either to the proximity effect with normal Mg metal or to the formation of nonstoichiometric $Mg_{1+x}B_2$.

(3) Effect of residual gas during growth: Residual oxygen is very harmful to the film growth of $MgB_2$ even with a $P_{O2}$ of as low as $1\times10^{-9}$ Torr. Residual hydrogen and nitrogen have a negligible or even a slightly favorable effect on $MgB_2$ film growth. The introduction of $H_2$ and $N_2$ gas improved $T_c$ by about 1 K although the origin of this slight improvement is unclear.

(4) Effect of in-situ annealing: The $T_c$ increased with increasing annealing temperature. The maximum improvement of ~ 2 K was achieved by annealing at 503°C. XRD and AFM measurements suggest that this improvement in superconductivity may result from the partial elimination of an unfavorable feature in the grain boundaries.

(5) Thickness dependence: $T_c$ decreased and the resistivity increased with decreasing film thickness in both as-grown and in-situ annealed films. A comparison of the physical properties of as-grown and in-situ annealed films showed that the latter have a dead surface (nonsuperconducting) layer and are unsuitable for fabricating junctions.

(6) Effect of substrates: A weak but finite substrate effect was observed. A hexagonal surface slightly improves the quality of the $MgB_2$ film.

These results provide much information related to the preparation of high quality superconducting $MgB_2$ films for such applications as Josephson junctions, superconducting wire, and coated conductor.

**Acknowledgements**

The authors thank Dr. S. Karimoto, and Dr. H. Yamamoto for fruitful discussions, Mr. N. Honma for ICP analysis, and Dr. H. Takayanagi and Dr. S. Ishihara for their support and encouragement throughout the course of this study.

**Figure captions**

Fig. 1: (A) Mg vapor pressure curve, (B) thermodynamic phase stability line of $MgB_2$ (from ref. 17), and (C) kinetically limited phase stability line of $MgB_2$ (from ref. 18). The dotted line ($10^{-4}$ Torr) is an upper limit of Mg vapor pressure compatible to physical vapor deposition.

Fig. 2: (a) Molar ratio of Mg to $B_2$ evaluated by ICP analysis in films grown at temperatures of 200°C to 500°C and also with Mg flux rates 1.3 to 10 times the nominal rate. (b) The superconducting transition temperature ($T_c$) of $MgB_2$ films grown at different substrate temperatures ($T_s$) on sapphire C with an Mg rate twice the nominal rate.

Fig. 3: (a) XRD pattern and (b) ρ-T curve of a typical as-grown $MgB_2$ film prepared on a sapphire C substrate. Peaks labeled "sub." correspond to substrate peaks in the XRD pattern. The inset in Fig. 3 (a) and 3 (b) shows a RHEED pattern and an enlarged view of the superconducting transitions for the film, respectively.

Fig. 4: (a) $T_c$ versus Mg flux ratio and (b) ρ-T curves of $MgB_2$ films prepared with different Mg fluxes at a fixed substrate temperature (280∘C).

Fig. 5: (a) XRD patterns and (b) ρ-T curves of $MgB_2$ films prepared in various $O_2$ pressures. The inset in figure (b) is an enlarged view of the superconducting transitions.

Fig. 6: Superconducting transitions of $MgB_2$ films prepared in various (a) $H_2$ and (b) $N_2$ pressures.

Fig. 7: $N_2$ and $H_2$ pressure dependence of $T_c$ of $MgB_2$ films. "$N_2$-1" and "$N_2$-2" indicate two different rounds of films prepared in various $N_2$ pressures.

Fig. 8: XRD patterns of $MgB_2$ films prepared in various $N_2$ pressures.

Fig. 9: (a) XRD patterns and (b) superconducting transitions of $MgB_2$ films annealed at various temperatures (380 ~ 526°C). The results of as-grown films are shown for comparison.

Fig. 10: AFM images (1 mm×1 mm) of an as-grown film (a) and post-annealed films at 480°C (b) and 526°C (c).

Fig. 11: (a) ρ-T curves of as-grown films with various thicknesses (50-1000Å) and (b) an enlargement view of the superconducting transitions.

Fig. 12: (a) ρ-T curves of post-annealed films (480°C for 10 min.) with various thicknesses (100-1000Å) and

(b) an enlargement view of the superconducting transitions.

Fig. 13: Thickness dependence of $T_c$ of as-grown and post-annealed films.

Fig. 14: RHEED patterns of the films with thickness of 50Å (a) before (as-grown) and (b) after post-annealing at 480°C for 10 minutes.

Fig. 15: Resistivity-versus-temperature curves of (a) as-grown and (b) post-annealed films (480°C for 10 min) around $T_c$ prepared on various substrates (Si (111) (filled circles), sapphire C (open circles), sapphire R (open squares), $SrTiO_3$ (100) (open diamonds)).

Fig. 16: XRD pattern (a) and ρ-T curve (b) of the $MgB_2$ film prepared on a glass substrate (Corning #7059). The inset in (b) is an enlarged view of the superconducting transition.

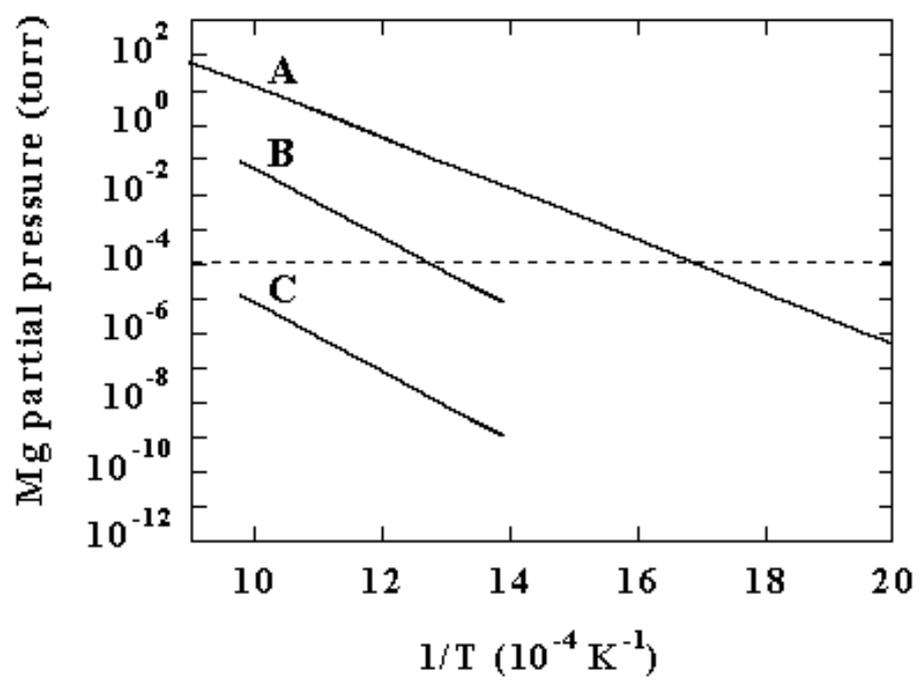

Fig. 1

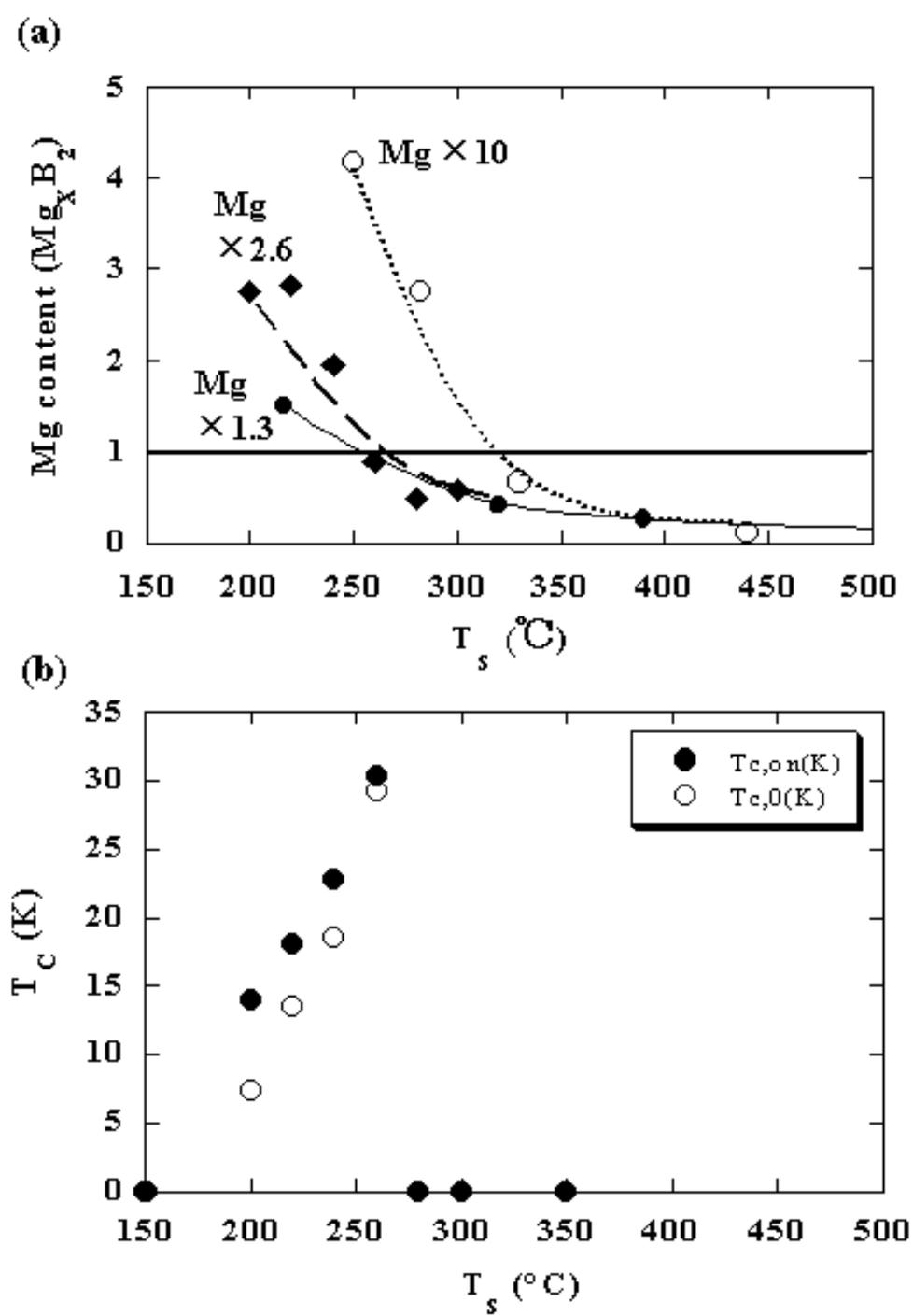

Fig. 2

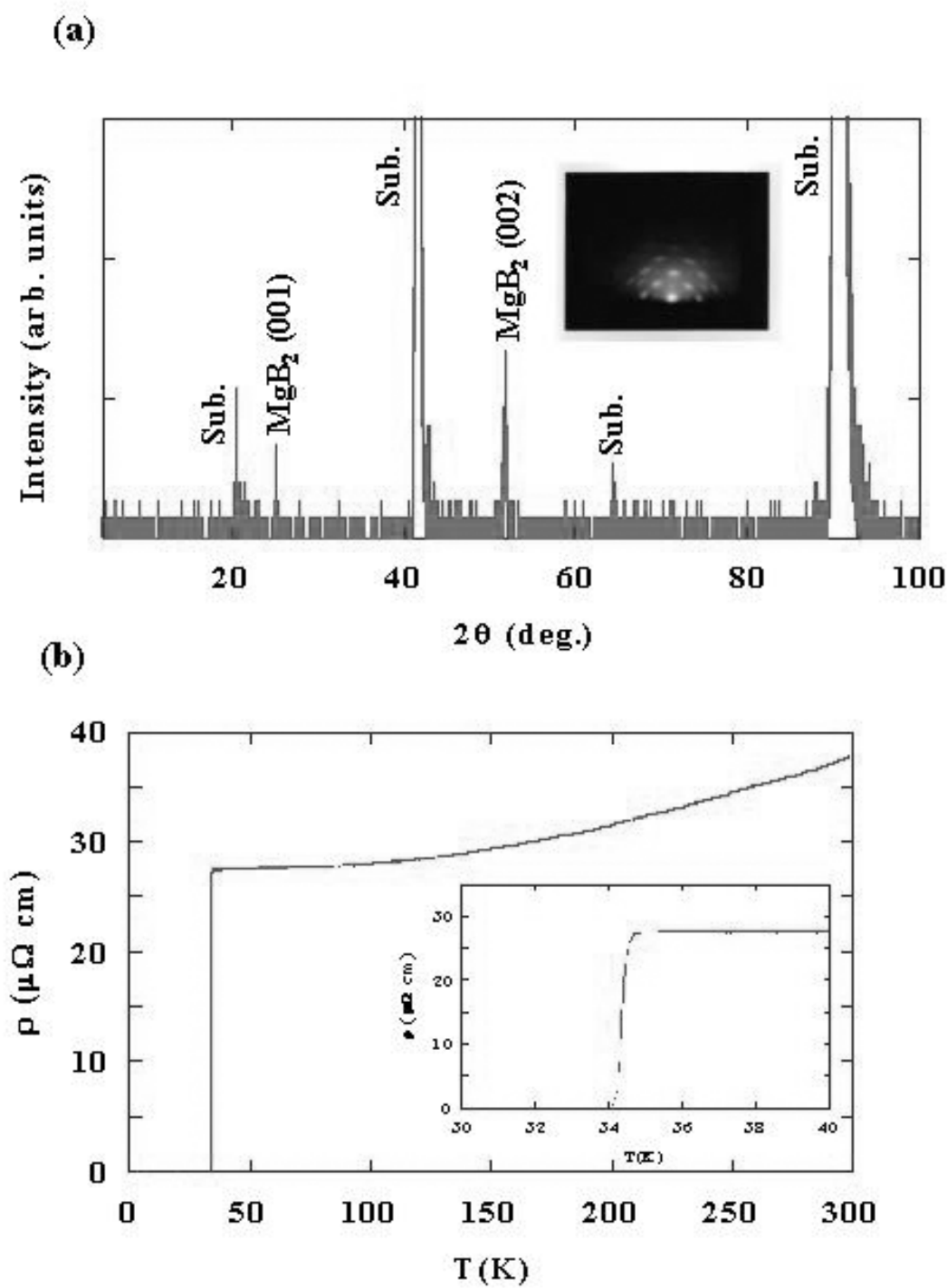

Fig. 3

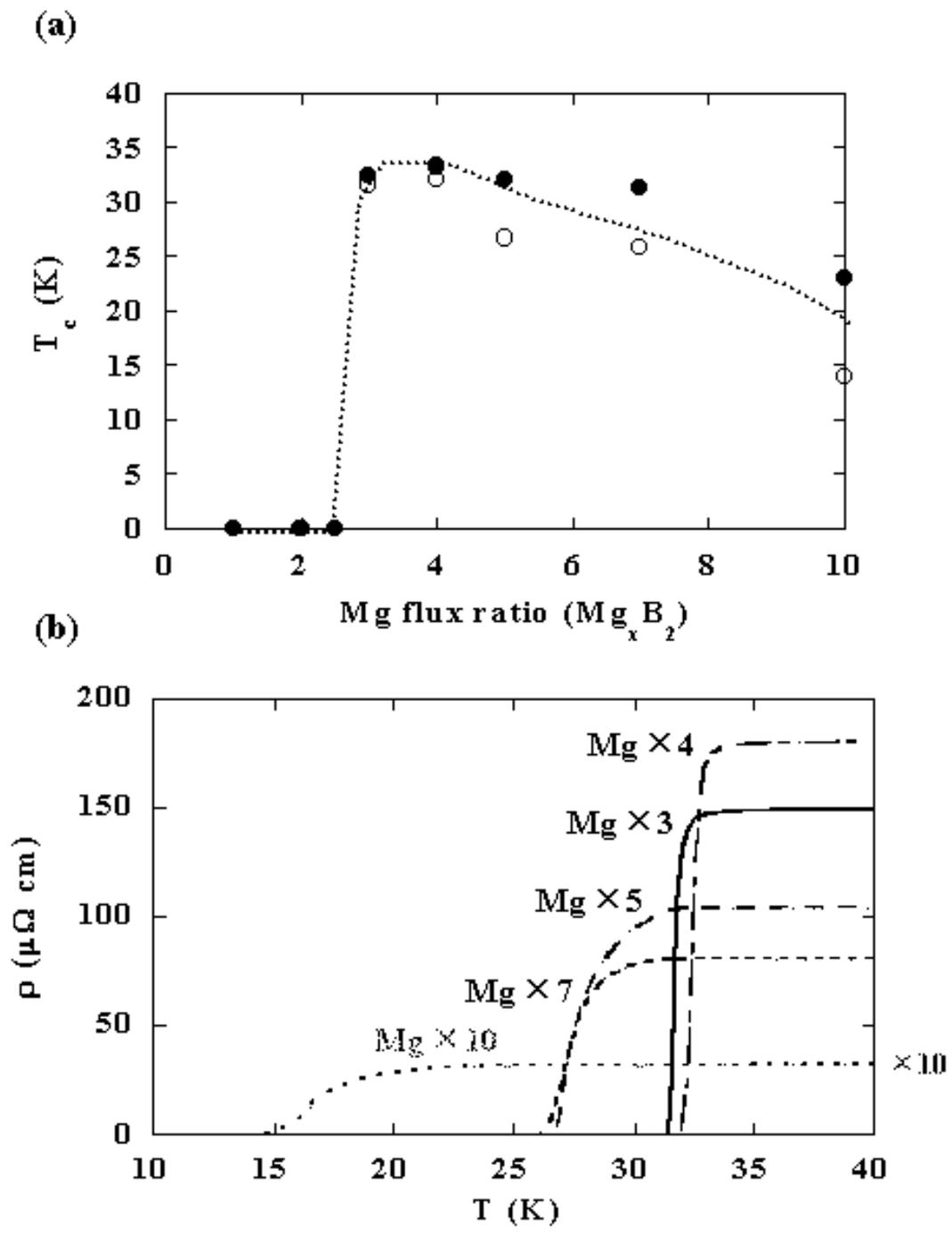

Fig. 4

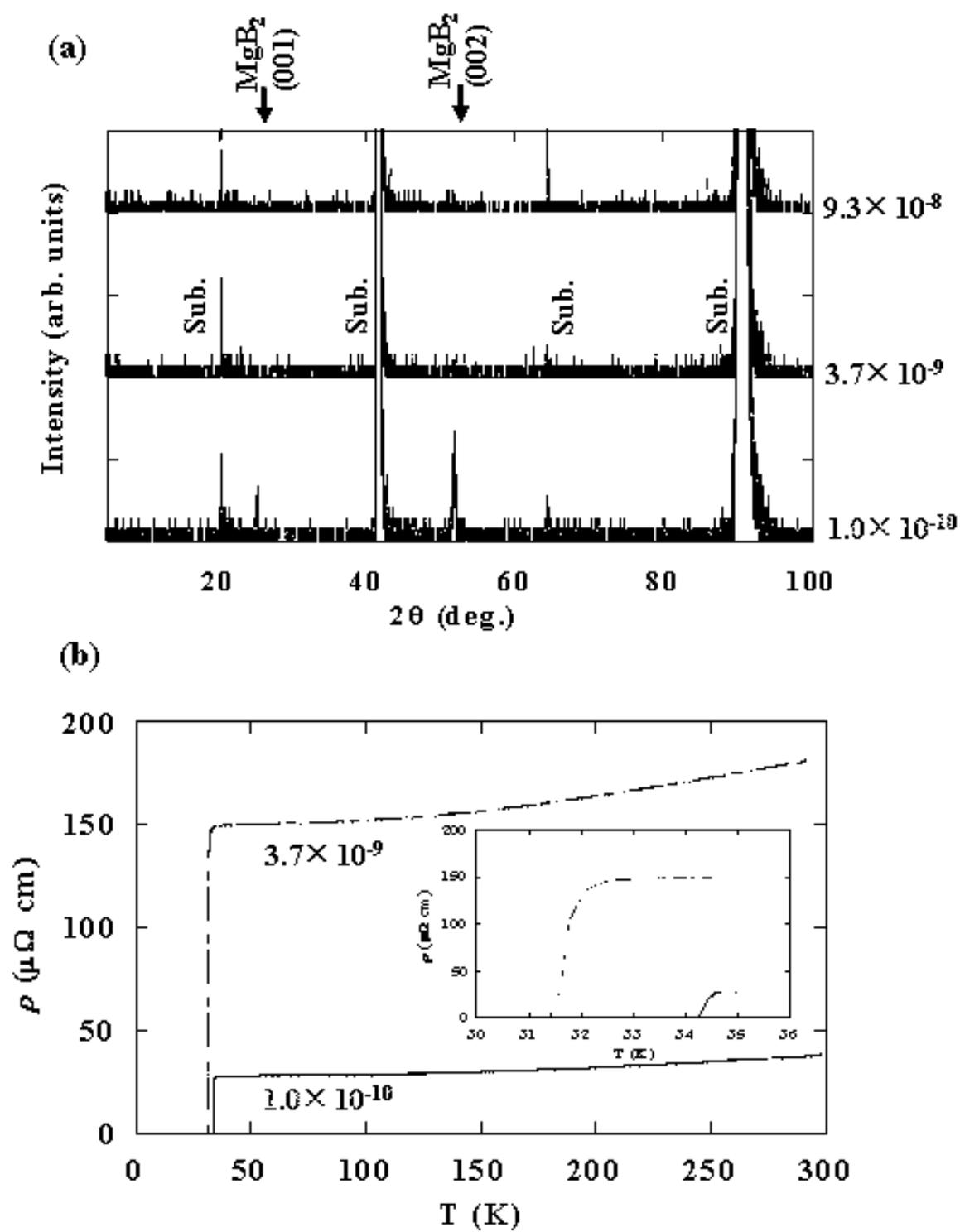

Fig. 5

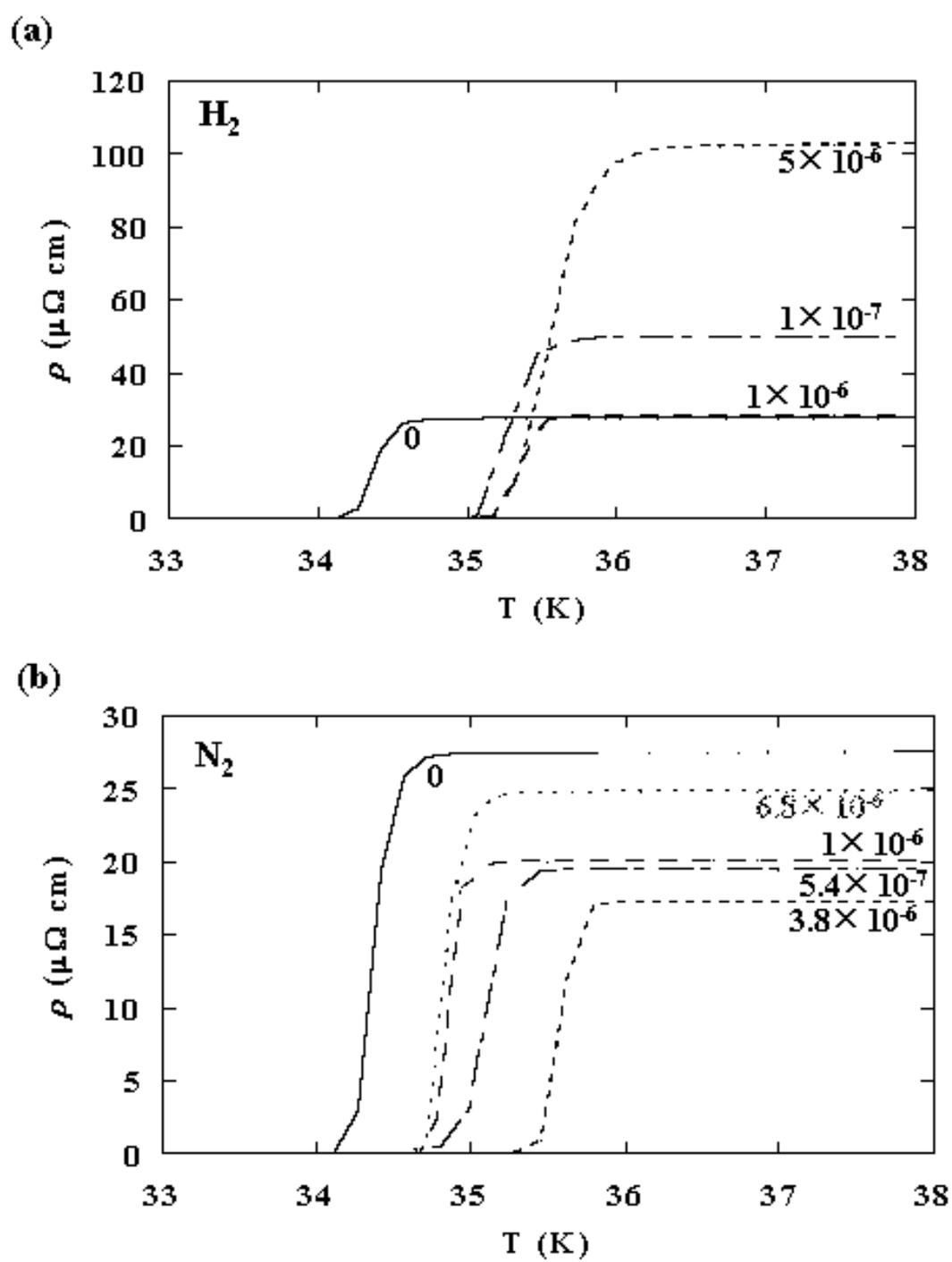

Fig. 6

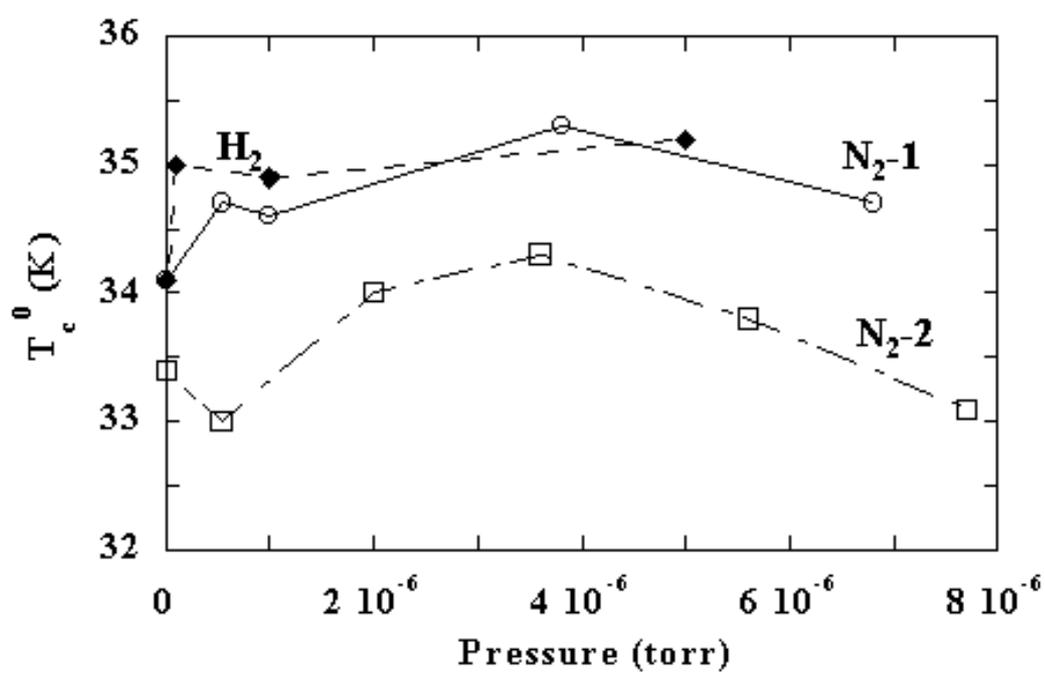

Fig. 7

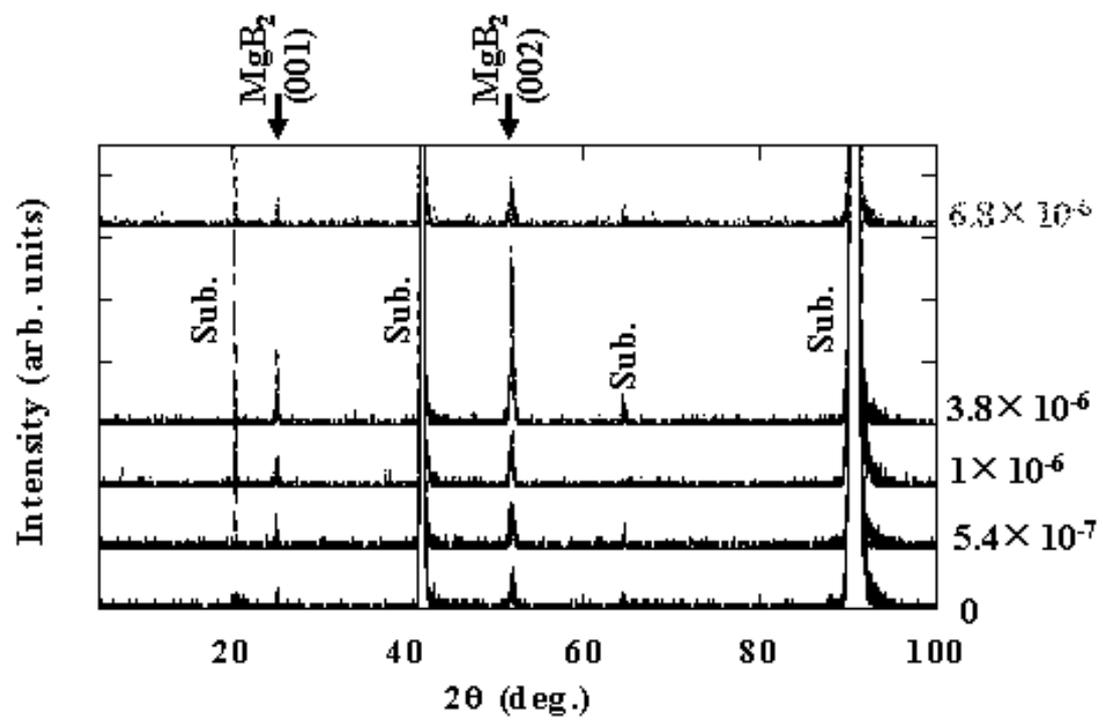

Fig. 8

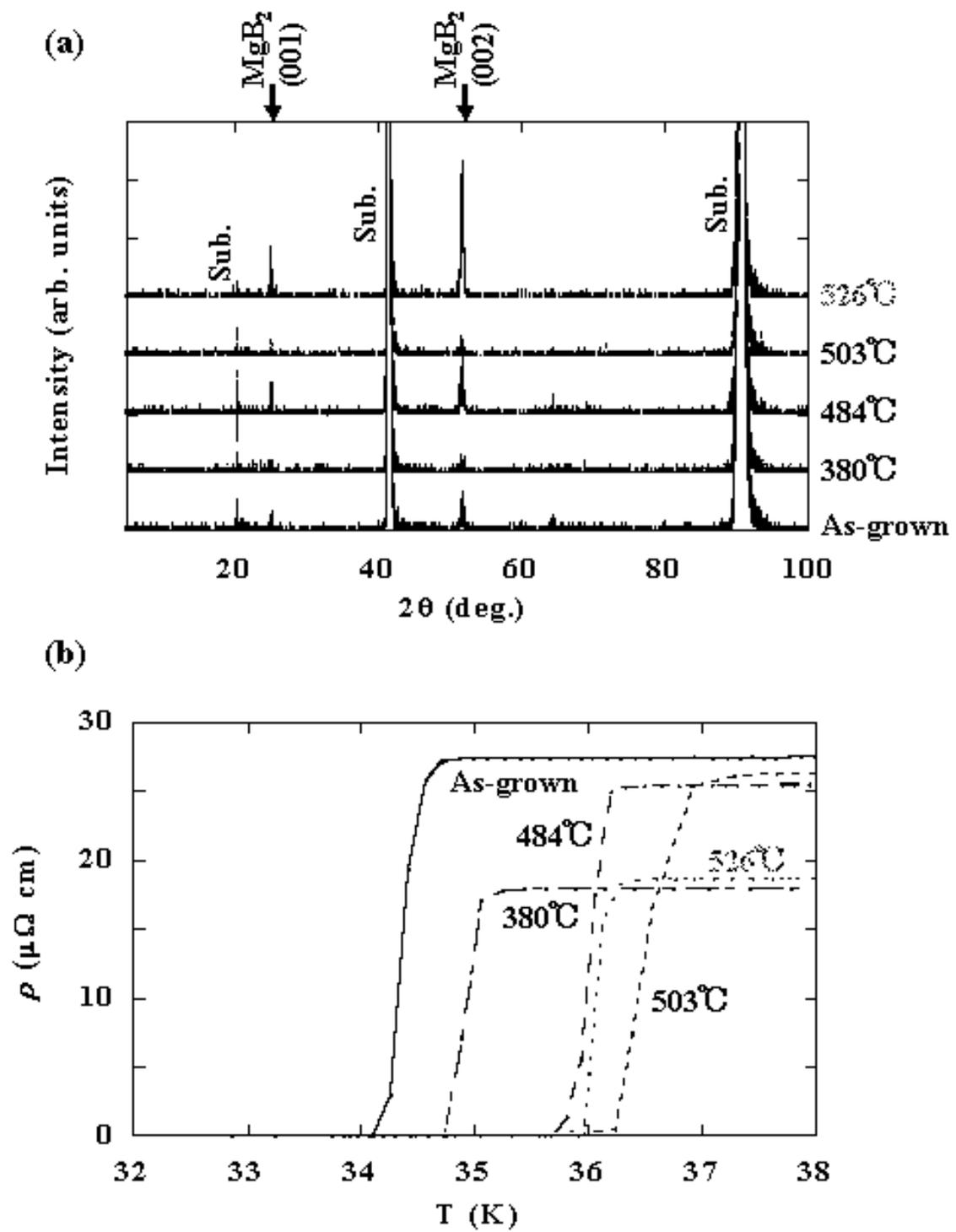

Fig. 9

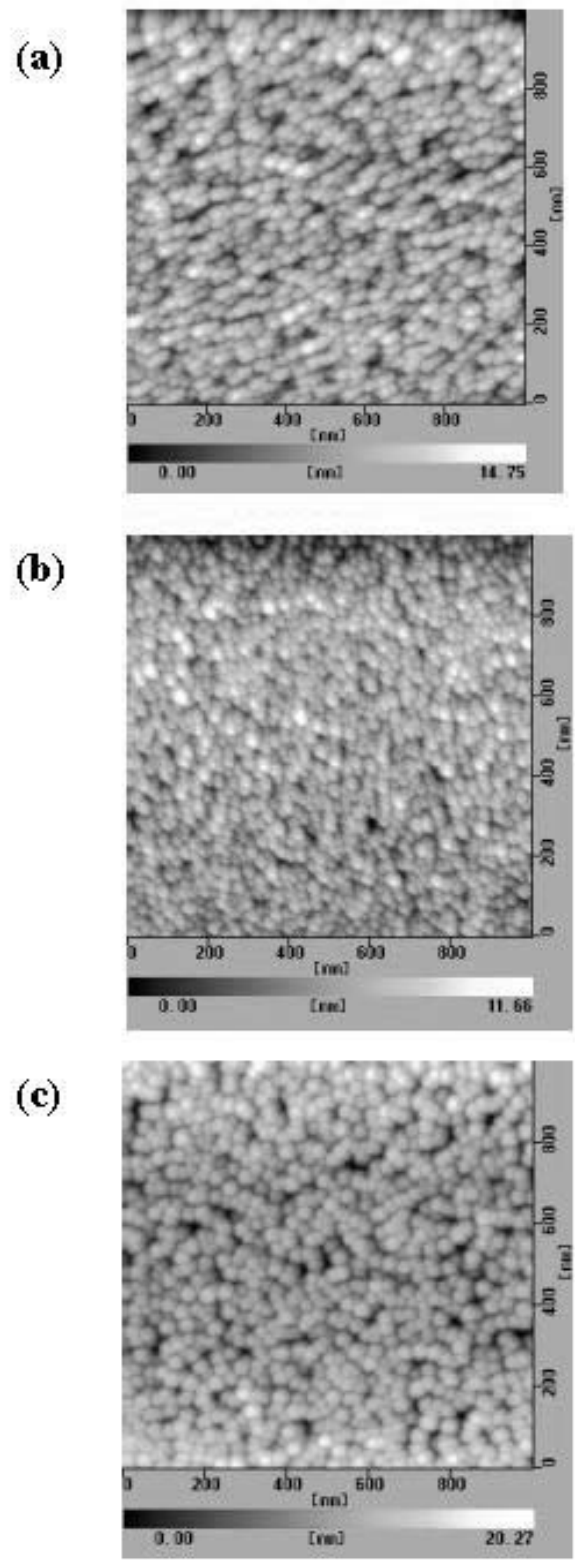

**Fig. 10**

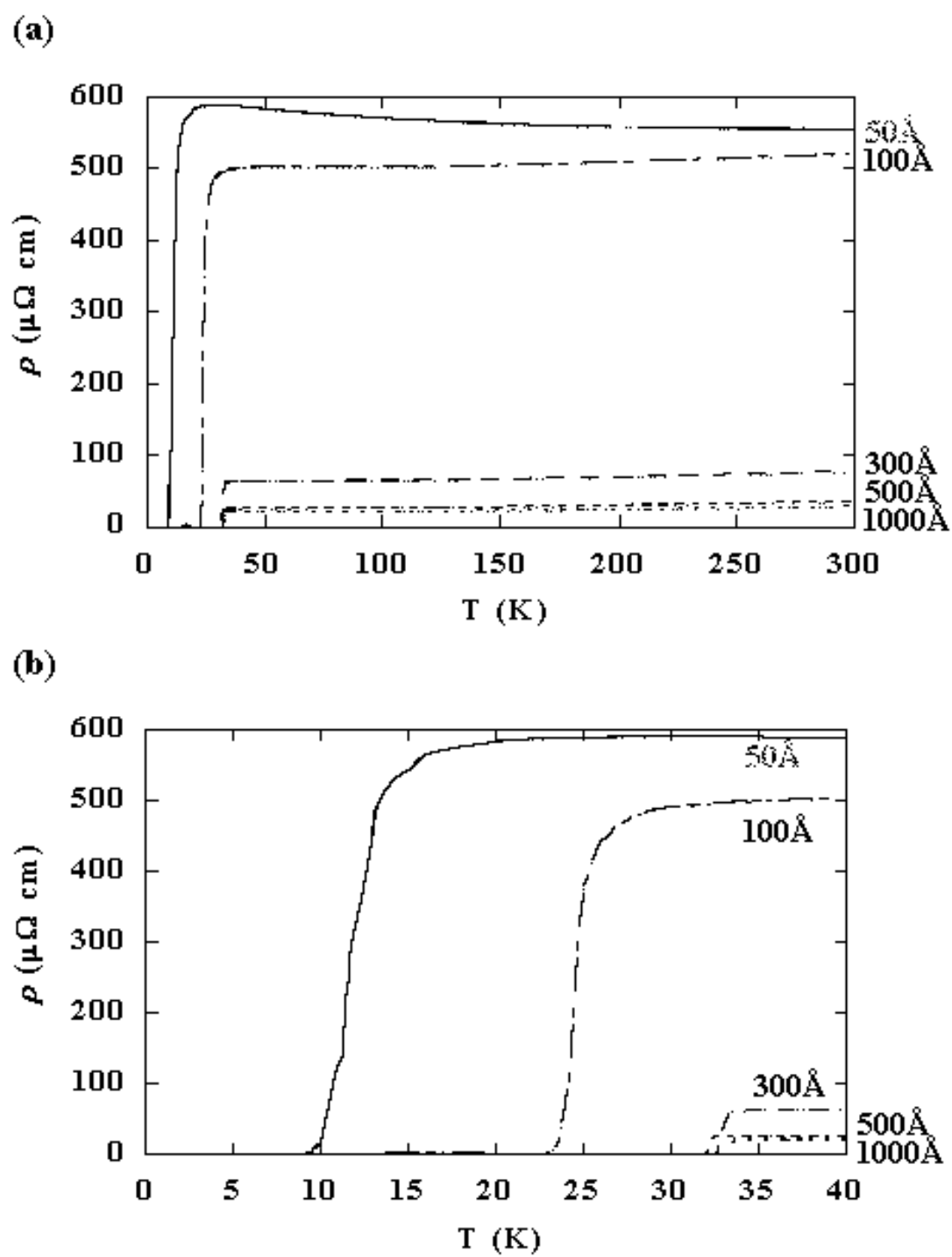

**Fig. 11**

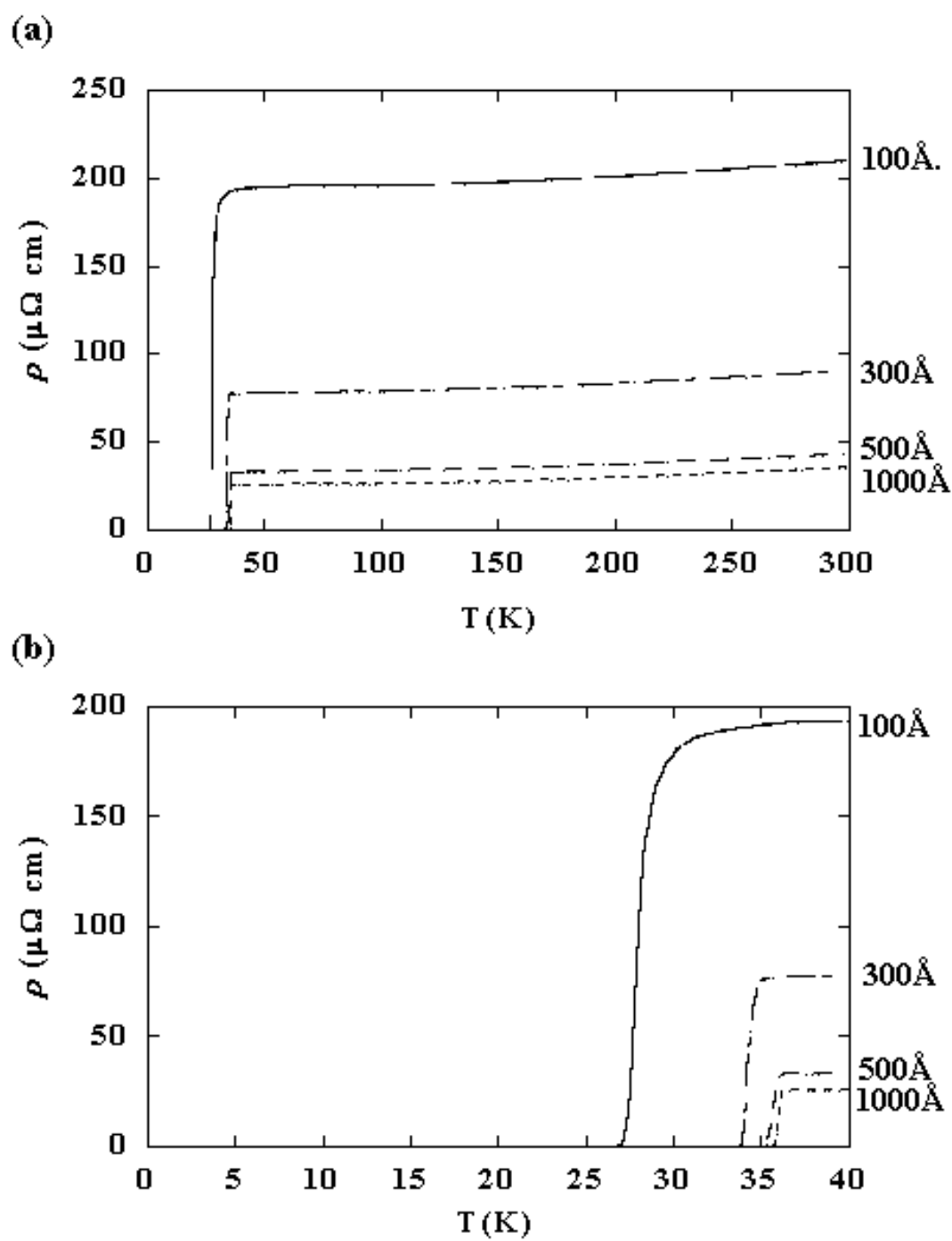

Fig. 12

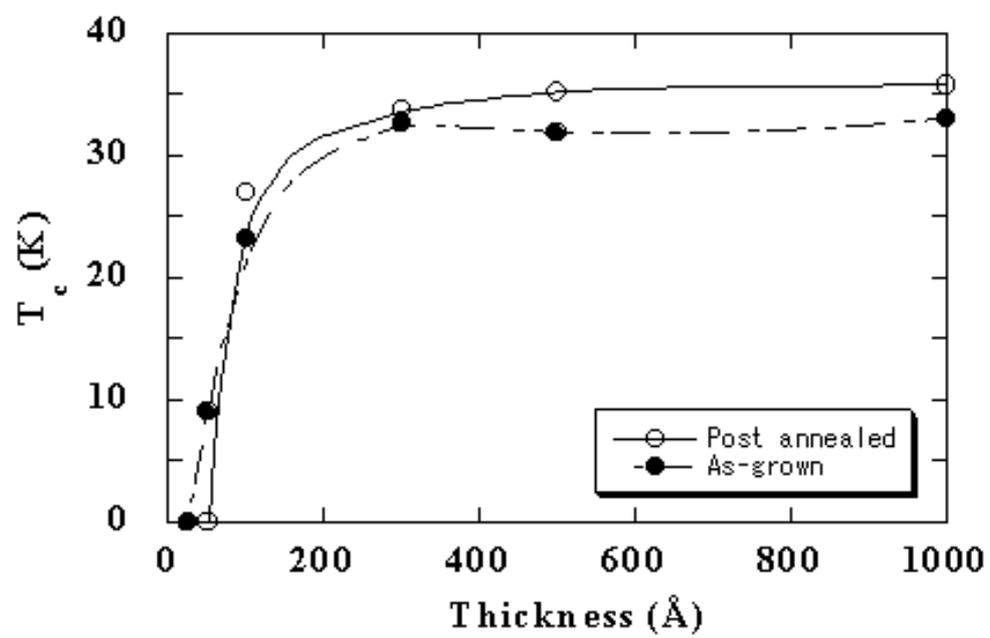

**Fig. 13**

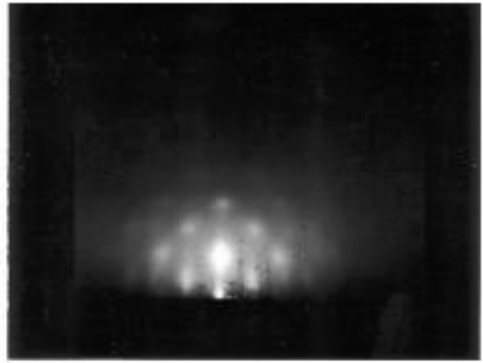

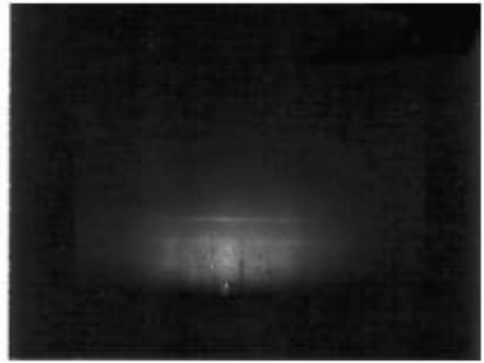

**Fig. 14**

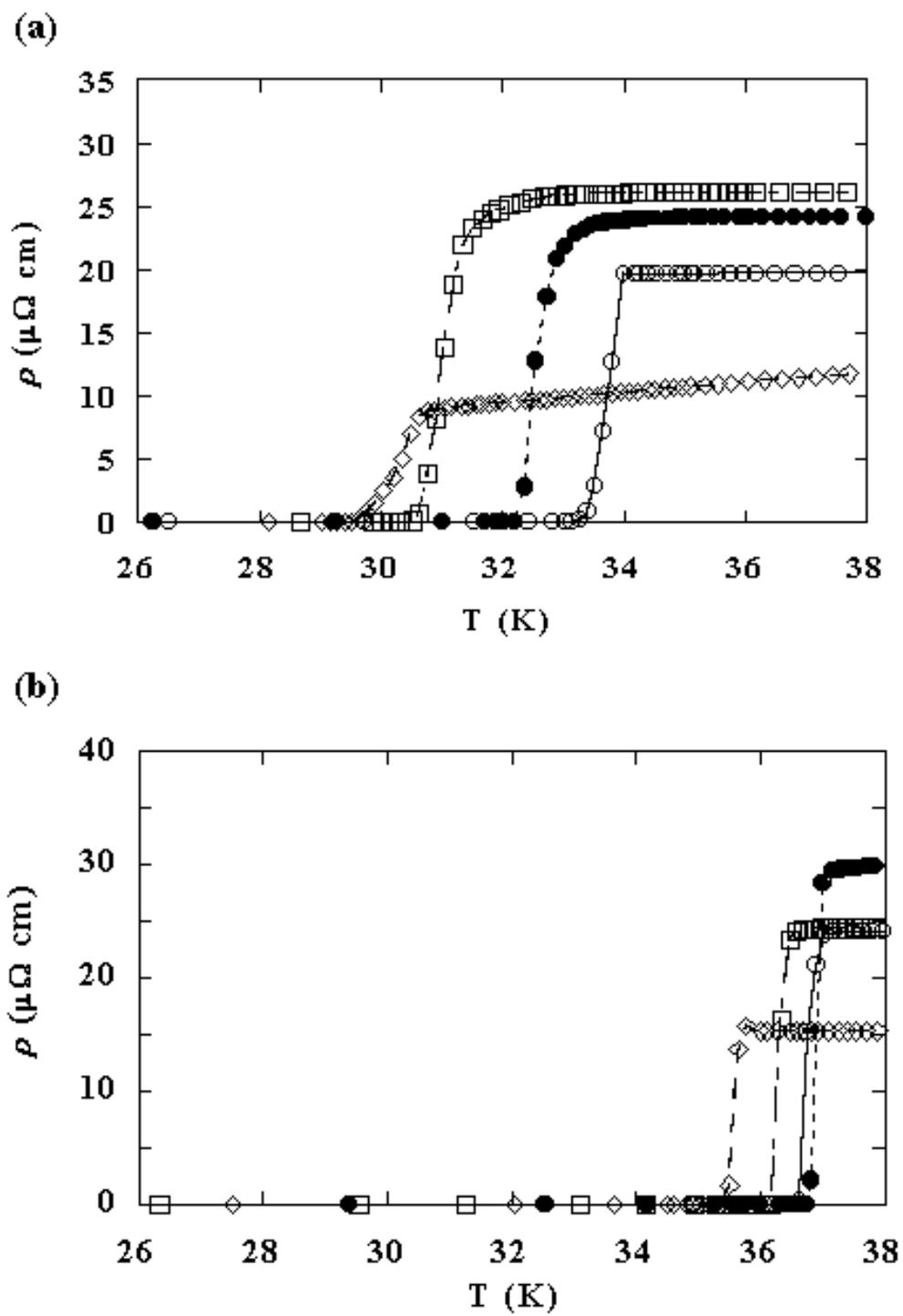

**Fig. 15**

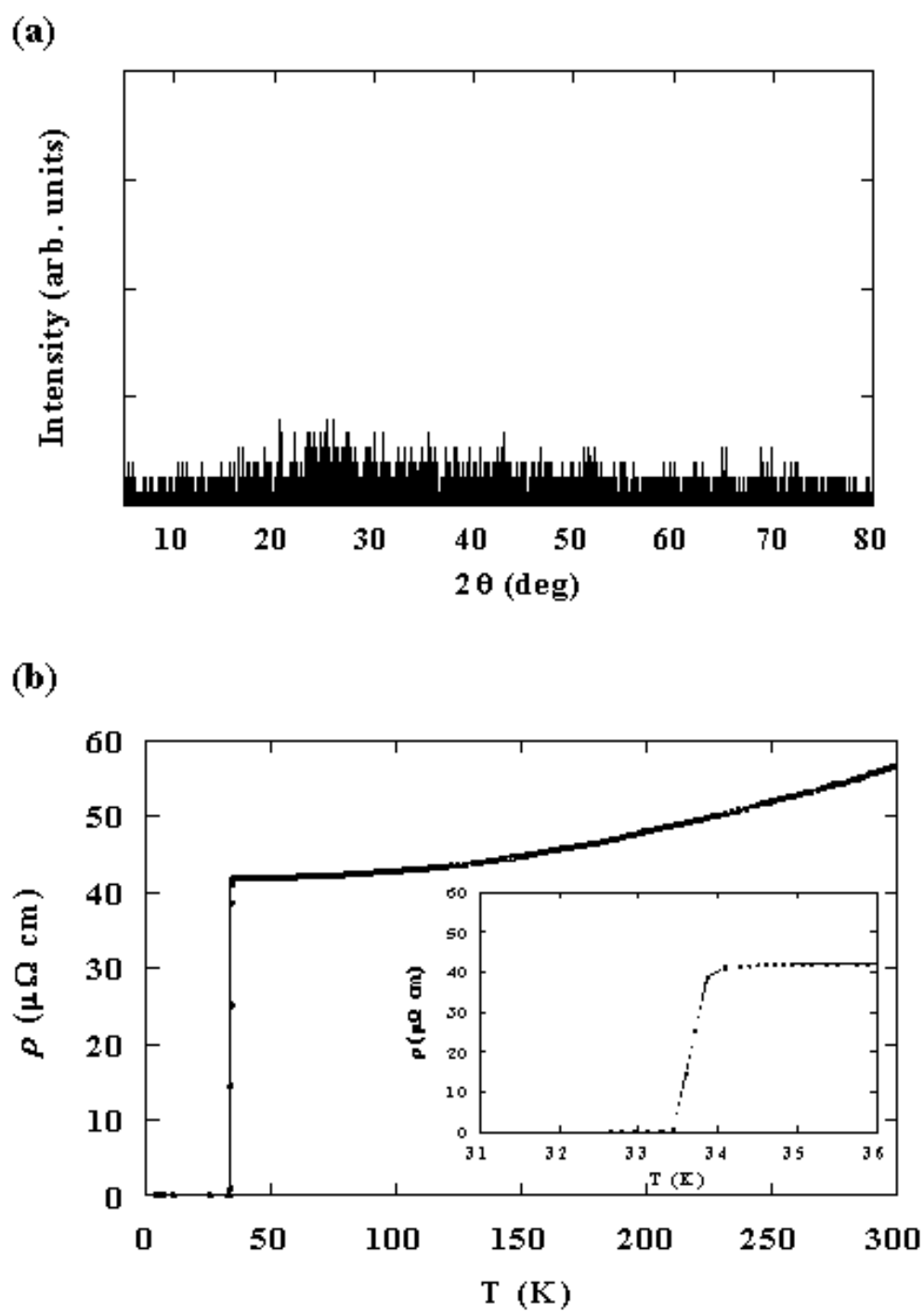

Fig. 16